# *Ab-Initio* computations of electronic and transport properties of wurtzite aluminum nitride (w-AlN)

Ifeanyi H. Nwigboji,[1] John I. Ejembi,[1] Yuriy Malozovsky,[1] Bethuel Khamala,[1] Lashounda Franklin,[1] Guanglin Zhao,[1] Chinedu E. Ekuma,[2] and Diola Bagayoko[1*]

[1]Department of Physics, Southern University and A&M College, Baton Rouge, Louisiana, 70813, USA.

[2]Department of Physics & Astronomy and Center for Computation and Technology, Louisiana State University, Baton Rouge, Louisiana, 70803, USA.

All Correspondence:

Diola Bagayoko

Phone: +1225 205 7482

Email: bagayoko@aol.com

Postal Address: Department of Physics, William James Hall, Southern University and A&M College, P. O. Box 11776, Baton Rouge, Louisiana, 70813, USA

## Abstract

We report findings from several *ab-initio*, self-consistent calculations of electronic and transport properties of wurtzite aluminum nitride (w-AlN). Our calculations utilized a local density approximation (LDA) potential and the linear combination of Gaussian orbitals (LCGO). Unlike some other density functional theory (DFT) calculations, we employed the Bagayoko, Zhao, and Williams' method, enhanced by Ekuma and Franklin (BZW-EF). The BZW-EF method verifiably leads to the minima of the occupied energies; these minima, the low laying unoccupied energies, and related wave functions provide the most variationally and physically valid density functional theory (DFT) description of the ground states of materials under study. With multiple oxidation states of Al ($Al^{3+}$ to Al) and the availability of $N^{3-}$ to N, the BZW-EF method required several sets of self-consistent calculations with different ionic species as input. The binding energy for ($Al^{3+}$ & $N^{3-}$) as input was 1.5 eV larger in magnitude than those for other input choices; the results discussed here are those from the calculation that led to the absolute minima of the occupied energies with this input. Our calculated, direct band gap for w-AlN, at the Γ point, is 6.28 eV, in excellent agreement with the 6.28 eV experimental value at 5K. We discuss the bands, total and partial densities of states, and calculated, effective masses.

**This paper has been submitted to Journal for publication**

## 1. Introduction and Motivations

The important technological applications of w-AlN have been described by many authors. [1-21] Wurtzite AlN (w-AlN) exhibits piezoelectricity, high thermal conductivity, mechanical strength and low electron affinity. [1] Furthermore, high solubility with other III-V compounds, good mechanical, and thermal matching with substrate materials make w-AlN a promising material in the fabrication of optical, [22] ultraviolet (UV), optoelectronic, and high frequency electrostatic devices and sensors. [11]

Yim *et al.* [23] and Perry and Rutz [24] provided some of the earliest experimental results on the band gap of w-AlN. Yim and his group studied the epitaxially grown single-crystal AlN layers utilizing optical absorption measurements at room temperature. The samples had a thickness of 0.25mm. They found the band gap to be direct, with a value of 6.2 eV. [23] The study of Perry and Rutz determined the band gap of w-AlN by optical absorption edge of single-crystal samples of AlN prepared by a close-spaced vapor process. The direct band gaps were found to be 6.28 eV and 6.2 eV at low (5k), and room temperatures, respectively. [24] The single-crystal AlN samples had typical surface areas of approximately 1cm x 1cm and thicknesses from less than 1μm to several μm. Loughin and French determined the band gap of w-AlN by vacuum ultraviolet (VUV) optical measurement. [3] Their method emphasized the relationship among critical points, grouping them into sets that are representative of transitions between pairs of bands. The direct band gap was found to be 6.2 eV, for their single crystal grown by a modified Bridgeman technique; we could not determine the measurement temperature. [3]

Measurements of the band gap of w-AlN single- crystal, stoichiometric thin films grown epitaxially by reactive magnetron sputtering, led to a value of 6.2 eV with an experimental error of ± 0.2 eV. [1] This work was done at a temperature of $300^{0C}$ and the sample was 10 x 10 x 0.5 $mm^3$. Vispute *et al.* [25], in their work on high quality epitaxial aluminum nitride layers grown on sapphire substrates, by pulsed layer deposition at a temperature of $800^{0C}$, found the band gap to be 6.1 eV. We could not determine the sample thickness. Li and coworkers [26] reported a low temperature (10K) band gap of $6.11 \pm 0.01$ eV. This value is smaller than low and room temperature gaps noted above. Li et al. argued that absorption, transmission, and reflectance measurements actually miss the fundamental band gap and obtain the separation between the conduction band minimum (CBM) and bands that are immediately below the valence band maximum (VBM).The transition corresponding to the fundamental gap is reportedly not active for $\alpha$-polarization; this polarization occurs when the excitation light propagates in the direction of the c axis. Taking the applicable uncertainties into account, the measured band gaps of w-AlN are between 6.0 and 6.3 eV. It should be recalled that sample quality, thickness, and measurement temperature and pressure can explain this spread of 0 to 0.3 eV. With this understanding, experimental values of the band gap are in general agreement with each other. We examine below results from theoretical calculations of electronic properties of w-AlN, with emphasis on the band gap as obtained from first principle calculations.

In 1994, Christensen and Gorczyca [27] calculated the band gap of w-AlN and obtained a value of 4.78 eV. According to the authors, they utilized the simplest version of the linear muffin-tin orbital (LMTO) method and the atomic-sphere approximation (ASA). [27] Another theoretical work of the same authors [27] reported an energy gap of 4.52 eV. [6] Here, they utilized the full potential (FP-LMTO) scheme. The self-consistent calculations of Ching et al., [28] using the first principle linear combination of atomic orbitals (LCAO) and the Wigner interpolation method, gave a band gap value of 4.4 eV. The plane-wave pseudopotential (PW-PP) calculations of Wright and Nelson [29] produced a band gap of 4.41 eV. The LDA work of Rubio et al. [30] led to a band gap of 3.9 for w-AlN. Also, the first principle pseudopotential calculations of Miwa and Fukumoto [16] for the w- AlN yielded a direct band gap of 4.09 eV. Another first principle pseudopotential calculation of Marco Buongiorno [31] and his group resulted in a band gap of 4.44 eV. Dridi et al. [32] performed relativistic, full-potential linear augmented plane - wave (FP-LAPW) calculations, with an LDA potential, to obtain a gap of 4.4 eV for w-AlN.

Magnuson et al.,[1], in their work of 2009, produced a band gap of 3.96 eV. They employed the Wu-Cohen GGA (WC-GGA) potential in their calculations. Another GGA result is that of Cui et al. [33] who found a band gap of 4.21 eV. Band structure calculations by Persson et al., [21] with the GGA potential of Perdew, Burke, and Ernzerhof (PBE), produced a band gap of 4.20 eV and a gap of 4.22 eV when they employed the exchange correlation of Perdew and Wang (PW). These GGA results are somewhat smaller than several LDA ones. Table I below contains other theoretical results, including those, unlike the ones above, that employed fitting, DFT potentials including additional parameters beyond those in the standard LDA and GGA, and Green function and screened Coulomb (GW) approximation that is beyond DFT.

The 2009 Modified Becke and Johnson (MBJ) LDA calculations of Tran and Blaha [34] using full potential linearized augmented plane waves plus local orbitals [FP-(L)APW+lo] led to a band gap of 5.55 eV for w-AlN. A Hoyd-Scuseria-Ernzerhof (HSE) hybrid density functional potential was employed by Yan et al. [5] to obtain a calculated gap of 5.64 eV. The GW calculations of Rubio et al. [30] obtained a gap of 5.8 eV.

Rinke et al. [35] utilized and optimized effective potential (OEPx) for the exchange along with an LDA correlation (cLDA) to calculate a gap of 5.73 eV. By following this calculation with a GW one, they found a w-AlN band gap of 6.47 eV. While their DFT result underestimates the band gap, the GW finding overestimates it.

**Table I**: Illustrative discrepancies between calculated and experimentally measured band gaps ($E_g$, in eV) of w-AlN, emphasis are placed on ab-initio results.

| Computational Methods | Potentials (DFT and others) | $E_g$ (eV) |
|---|---|---|
| LTMO +ASA | LDA | 4.78 [a] |
| FP-LMTO | LDA | 4.52 [b] |
| pseudopotential Plane-wave | LDA | 4.41 [c] |
| Semi-relativistic pseudopotential | LDA | 3.9 [d] |
| LCAO + Wigner interpolation | LDA | 4.4 [e] |
| Pseudopotentials | LDA | 4.09 [f] |
| Pseudopotentials | LDA | 4.44 [g] |
| FP-LAPW | LDA | 4.4 [h] |
| PW | GGA | 4.22 [i] |
| PBE | GGA | 4.20 [i] |
| Pseudopotential plane-wave | GGA | 4.21 [j] |
| APW+lo | WC-GGA | 3.96 [k] |
| LMTO+ASA & combined correction terms | LDA [with correction] | 6.05 [a] |
| Exact-exchange pseudopotential | OEPx(cLDA)+$G_0W_0$ | 6.47 [l] |
| Exact-exchange pseudopotential | OEPx(cLDA) | 5.73 [l] |
| FP-(L)APW+lo | MBJLDA | 5.55 [m] |
| Plane wave +Hybrid density functional | HSE06 | 5.64 [n] |
| Pseudopotential | GGA + XC | 6.3 [o] |
| Plane-wave pseudopotential | GW | 5.8 [d] |
| Pseudopotential | Empirical pseudopotential model (EPM). (fitting) | 6.11 [p] |
| Pseudopotential | EPM | 6.22 [q] |
| Pseudopotential | Semi-empirical tight binding (fitting) | 6.2 [r] |
| **Experiment (low temperature) 5k** | Absorption Measurement. | **6.28** |
| **Experiment (Room temperature)** | Absorption Measurement | **6.2** |

[a] Ref. [27], [b] Ref. [6], [c] Ref. [29], [d] Ref. [30], [e] Ref. [28], [f] Ref. [16], [g] Ref. [31], [h] Ref. [32], [i] Ref. [21], [j] Ref. [33], [k] Ref. [1], [a] Ref. [27], [l] Ref. [35], [l] Ref.[35], [m] Ref. [34], [n] Ref. [5], [o] Ref. [36], [d] Ref.[30], [p] Ref.[37], [q] Ref.[39], [r] Ref.[38]

A pseudopotential calculation within the generalized gradient approximation for the exchange correlation (GGA), [36] followed by a GW calculation, led to a band gap of 6.3 eV. The empirical pseudopotential model (EPM) work of Fritsch *et al.* [37] reported a band gap of 6.11 eV for w-AlN, from both their anisotropic and isotropic pseudopotential calculations. Kobayashi *et al.* [38] found a gap of 6.2 eV with semi empirical tight-binding calculations. The EPM calculation of Rezaei [39] and his group resulted in a band gap of 6.22 eV. The fitting involved in EPM and tight binding calculations explain the apparent agreement of the resulting gaps with measured ones.

While the band gap of AlN is well established experimentally, at low and room temperatures, theoretical findings from ab initio DFT calculations seriously disagree with corresponding, measured ones. While non-DFT calculations, mainly GW ones show vast improvements, the resulting gaps still under or overestimate the experimental ones. We aim to contribute to the resolution of the above band gap problem, for w-AlN, by using a method that has led to accurate DFT (mainly LDA) description and prediction of properties of semiconductors.

## 2. Method and Computational details

In this work, we employed the Ceperley and Alder [40] local density approximation (LDA) potential as parameterized by Vosko and his group. [41] We utilized the linear combination of Gaussian orbital (LCGO) formalism. These features of our computations are the same as those of most of the other LDA calculations, except for the use of exponential functions by some at the place of Gaussian ones. Unlike other DFT calculations noted above, we employed the Bagayoko, Zhao, and Williams (BZW) method [42-46], as enhanced by Ekuma and Franklin [47-50] in carrying out our self-consistent calculations.

This robust method involves a basis set optimization, after starting with a clearly small basis set that is to be no smaller than the minimum basis set. The minimum basis set is the one just large enough to account for all the electrons on the atomic or ionic species in the material under study. The modified version of the method, BZW-EF, differs from the original one by the rule utilized in increasing the size of the basis set. The original method (BZW) added orbitals representing unoccupied states in the order of increasing energy in the atomic or ionic species. The improvement consists, for a given principal quantum number at any given site, of adding p, d and f orbitals, if applicable, before the spherically symmetric s orbital for that principal quantum number. As apparent in the results of the BZW-EF method, this change recognizes the fact that polarization has primacy over spherical symmetry as far as the valence electrons in multi-atomic materials are concerned [49, 50]

A summary description of the method follows. Upon selecting a small basis set to perform completely self-consistent calculations, the method requires a second calculation with a basis set consisting of the original one augmented by one orbital. Naturally, for s, p, d, and f orbitals, this addition means the increase of the size of the basis set by 2, 6, 10, and 14, respectively, taking the spin into account. The occupied energies from the two calculations are compared graphically and numerically. As expected, they are generally found to be different – with occupied energies from Calculation II being lower than corresponding ones from Calculation I. Upon augmenting the basis set of Calculation II, Calculation III is carried out and the resulting occupied energies are compared to those of Calculation II. This process continues until the occupied energies of three consecutive calculations are found to be the same, within our computational uncertainties of 5 meV. With such three calculations, minimization process is completed. Among these three calculations, the one with the smallest basis set is the one providing the DFT description of the materials under study. The basis set for this calculation is referred to as *the optimal basis set*. Three successive calculations are required, as opposed to just two, as done in the past, due to the fact that local minima do exist for some materials. We found this situation for some zinc blende materials and for wurtzite GaN.

The other two calculations above leading to the same occupied energies and that have basis sets larger than the optimal one are not selected for the simple reason that once the occupied energies reached their absolute minima, the physical content of the Hamiltonian, for the description of the ground state, is not modified by using larger basis sets that contain the optimal one. With these basis sets much larger than the optimal one, each of which contains the optimal basis set, the Rayleigh theorem [47, 48, 50, 51] explains the lowering of some unoccupied energies below their values obtained with the optimal basis set while the occupied energies do not change.

The requirement of the BZW-EF method to obtain the absolute (lowest) minima of the occupied energies implicitly implies other careful considerations for calculations beside increasing the basis set. Specifically, in the case of w-AlN, the BZW-EF method also requires the examination of the effect of different input species in light of the three positive oxidation states of Al and the availability of three negative oxidation states for N. Indeed, the method is clear about ascertaining that the absolute minima of the occupied energies are attained before any claim of DFT description can be made. Hence, given that iterations only entail a "linear" minimization in the LCAO formalism, through the changes in the expansion coefficients, their self-consistent outputs do not necessarily coincide with the ground state properties. So, for w-AlN, the method demands several sets of ab-initio calculations with different input "ionic" species, i.e., from $Al^{3+}$ and $N^{3-}$, $Al^{2+}$ and $N^{2+}$, to $Al^{1+}$ and $N^{1-}$ or neutral Al and N. For this reason, this work entailed several sets of self-consistent calculations of electronic properties of w-AlN using three different ionic configurations as shown below, in the Section on results.

The program package we used in these calculations is from the Ames laboratory of the US department of Energy (DOE), Ames, Iowa. [52] The calculations are non-relativistic and are performed at a low temperature lattice constant. We start with the LCAO, self-consistent calculations of the electronic energy levels of the atomic or ionic species present in the system under study.

We provide below computational details pertinent to a replication of our work. The experimental [23], low temperature wurtzite lattice parameters used in our calculations are a = 3.112 Å and c =4.982 Å, with a "u" parameter of 0.382. For each of the solid state calculations with different ionic species as input, self-consistent calculations of electronic properties of said ions provided the input orbitals utilized to construct the solid state wave function in the LCGO formalism. In the construction of the atomic orbitals, we used even-tempered Gaussian functions, where the s and p orbitals of Al were described with 18 even-tempered Gaussian functions. The minimum and maximum exponents of these functions are 0.13 and $0.356 \times 10^5$, respectively. The same number of Gaussian functions was used for s and p orbitals for N, but with the minimum and maximum exponents of 0.12 and $0.160 \times 10^5$, respectively. In the iterative process, we utilized a mesh of 24 k- points in the irreducible Brillouin zone. Self-consistency was partly determined, in solid state calculations, by ensuring a change of no more than $10^{-5}$ [43] in the potential for the last iteration as compared to the one immediately preceding it. Self-consistency for the computations was reached after about 60 iterations. The computational error for the valence electrons was 0.00187 for the 32 electrons, or $5.8 \times 10^{-5}$ per electron.

## 3. Results.

Table II shows the successive, self-consistent calculations we performed, with $Al^{3+}$ and $N^{3-}$ as input species in our search for the optimal basis set of the BZW-EF method. It is interesting to note that for w-AlN, these calculations do not produce gaps of the same nature, i.e., direct or indirect. The binding energy for the above input ions is -16.86, over 1.5 eV larger in absolute value than the binding energies obtained with $Al^{2+}$ and $N^{2-}$ and with $Al^{1+}$ and $N^{1-}$. Additionally, as per the content of Table III, the width of the group of upper valence bands is 7.23 eV for $Al^{3+}$ and $N^{3-}$ as input while it is just 6.2 and 6.0 for $Al^{2+}$ and $N^{2-}$ and $Al^{1+}$ and $N^{1-}$, respectively. In light of the variational derivation of DFT, its description of w-AlN is given by the calculation that produces the absolute (lowest) minima of the occupied energies. Hence, the optimal basis set calculation for the input ions of $Al^{3+}$ and $N^{3-}$, i.e., Calculation III in Table II, provides the LDA description of w-AlN. The results from this calculation are the ones discussed below for w-AlN.

The first major point about our results consists of the excellent agreement between the calculated, direct band gap (6.28 eV) and the low temperature (5K) measured value of 6.28 eV reported by Perry and Rutz. Other features of the LDA description of the electronic structure of AlN are apparent in Figures 1, 2, and 3 for the band structure, the total density of states (DOS), and the partial densities of states (pDOS), respectively.

**Table II**. Successive calculations of the BZW-EF method for wurtzite AlN, with $Al^{3+}$ and $N^{3-}$ input "ionic" species. Calculation III led to the absolute minima of the occupied energies. ML below signifies that the conduction band minimum is between the M and L symmetry points. It is much closer to L, however.

| Calculations | Valence Orbitals for $Al^{3+}$ | Valence Orbitals for $N^{3-}$ | Number of Functions | Band Gaps (eV) |
|---|---|---|---|---|
| I | $2s^2\ 2p^6$ | $2s^2\ 2p^6$ | 32 | N/A |
| II | $2s^2\ 2p^6\ 3s^0$ | $2s^2\ 2p^6$ | 36 | 3.591 $\Gamma$-$\Gamma$ |
| III | $2s^2\ 2p^6\ 3s^0\ 3p^0$ | $2s^2\ 2p^6$ | 48 | **6.278** $\Gamma$-$\Gamma$ |
| IV | $2s^2\ 2p^6\ 3s^0\ 3p^0\ 3d^0$ | $2s^2\ 2p^6$ | 68 | 6.114 $\Gamma$-K |
| V | $2s^2 2p^6 3s^0 3p^0 3d^0 4p^0$ | $2s^2\ 2p^6$ | 74 | 5.052 $\Gamma$-ML |

**Table III**. Calculated band gap, its direct or indirect nature of the gap, upper valence band width, and the total valence band width of w-AlN, depending on the input "ionic" species. These numbers are obtained from the calculations that led to the minima of the occupied energies for the various input choices, as per the BZW-EF method. ML in Column 3 means that the conduction band minimum is between M and L symmetry points.

| Input Species | Band Gap | Nature of the Band Gap | Upper valence band width | Total Valence Band width |
|---|---|---|---|---|
| $Al^{+1}\ N^{-1}$ | 5.093 eV | $\Gamma$-ML, indirect | 6 eV | 15.20 eV |
| $Al^{+2}\ N^{-2}$ | 5.197 eV | $\Gamma$-ML, indirect | 6.2 eV | 15.30 eV |
| $Al^{+3}\ N^{-3}$ | 6.278 eV | $\Gamma$-$\Gamma$, direct | 7.23 eV | 16.85 eV |

Table IV lists the valence and low laying conduction band energies at high symmetry points. We could not find experimental reports that permit a detailed comparison of these bands and specific energies with corresponding, measured values. It is expected, however, that the content of Table IV will enable comparisons of our findings with future experimental results from optical absorption, photoemission, and various x-ray spectroscopy measurements.

The total width of the valence band is 16.89 eV, as given in Table IV and apparent in Figure I. Figure I shows the two groups of lower and upper valence bands whose respective widths can be derived from the content of Table IV as 2.792 eV and 7.235 eV, respectively. Without any broadening, the total density of the valence states exhibits peaks at -2.12, -6.6, and -14.7 eV. For the low laying conduction states, the DOS peaks are estimated at 9.17, 11.96, 12.65, and 13.73 eV. While the peaks above 10 eV may not be very reliable, the BZW-EF method gives low-laying, unoccupied energies in agreement with experimental findings. From Fig. 3 for the pDOS, one sees that the upper valence states are clearly dominated by nitrogen p (N-p) hybridized with a much smaller contribution from aluminum p (Al-p). The lowest laying group of valence bands is mostly from N-s, with a little contribution from Al-p.

While N states clearly dominate in the valence bands, Al-p and Al-s mostly make up the low laying conduction bands.

Even though we do not know of several experimental results for detailed comparisons, some of our results are in striking agreement with measurements by Magnuson et al. [1] Indeed, these authors reported N-2s – Al-3p and N-2s – Al-3s hybridization at the bottom of the valence band; this result is apparent in our Fig. 3, even though the contribution from Al -s is barely noticeable. Further, they placed this hybridization around -13.5 to -15 eV. Our total DOS peak at -14.7 agrees with this result. They stated that Al-3s and N-2s hybridization is observed at -15 eV. From the pDOS plots in Fig. 3, N-s and the tiny contribution from Al-s are indeed around -15 eV. They located a shoulder at -5 to -5.5 eV that is reportedly a signature of additional hybridization of N-2p and Al-3p; our DOS in Fig. 2 clearly shows this shoulder between -5 and -5.5 eV. The hybridization of the bonding N-2p with Al-2p, between -1 eV and -5 eV, as shown in our Fig.3, is well noted by Magnuson et al. [1] In our Fig. 3 for pDOS, while N-p extends down to around -7 eV, Al-p practically vanishes below -5 eV, leading to yet another agreement with detailed experimental findings.

The effective mass is a measure of the curvature of the bands. An agreement between measured and calculated effective masses clearly indicates the accuracy of the shape and curvature of the affected, calculated bands. [50] Our calculated electron effective masses around the Γ point are 0.303, 0.370, and 0.245 in the Γ-M, Γ-K, and Γ-A directions, respectively. Our calculated electron effective masses are in a general agreement with the calculated values of 0.27, 0.33, and 0.25 by Masakatsu [8] and his group. They are also in agreement with the 0.30 and 0.32 results of Persson [21] and his group. Experimental results are needed to verify these values, but to the best of our knowledge; there are no experimental results available on electron effective masses of w-AlN.

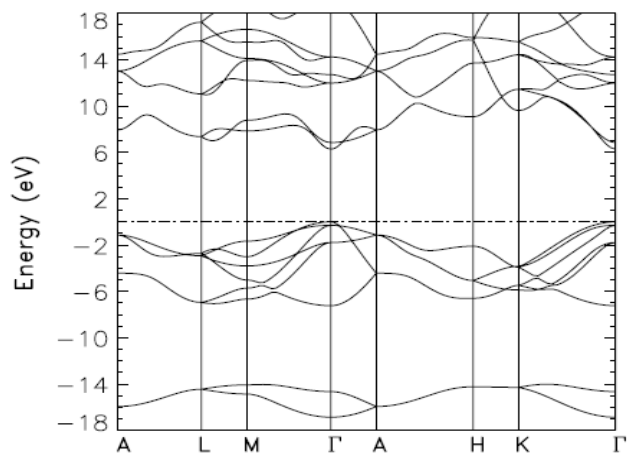

**Figure 1**. The calculated band structure (solid lines) of w-AlN as obtained with the optimal basis set of the BZW-EF method in Calculation III, with $Al^{3+}$ and $N^{3-}$ as input species. The low temperature lattice constants are a= 3.112 Å and c= 4.982 Å, with u=0.382.The dotted, horizontal line indicates the position of the Fermi energy that has been set equal to zero.

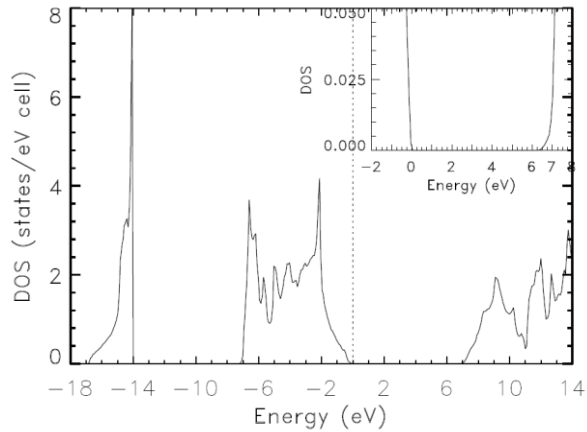

**Figure 2**. Calculated, total density of state (DOS) for w-AlN, as derived from the bands shown in Fig. 1, from Calculation III. The dotted, vertical line indicates the position of the Fermi energy that has been set to zero.

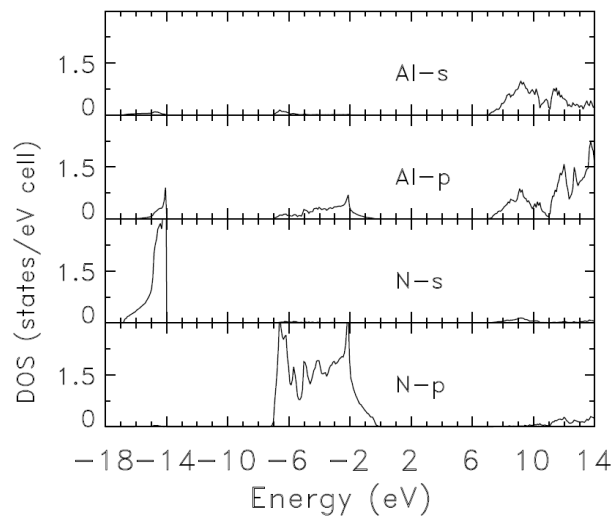

**Figure 3**. Partial density of states (pDOS) of w-AlN, as obtained from the bands shown in figure I.

**Table IV.** Calculated, electronic energies (in eV) of w-AlN at high symmetry points in the Brillouin zone. These energies are obtained from calculation III. The Fermi energy is set equal to zero. The calculated direct band gap is 6.28 eV.

| A | L | M | Γ | H | K |
|---|---|---|---|---|---|
| 13.027 | 15.619 | 14.106 | 12.706 | 15.283 | 14.424 |
| 13.027 | 11.040 | 13.927 | 11.979 | 13.714 | 14.424 |
| 13.027 | 11.040 | 12.208 | 11.979 | 13.714 | 14.424 |
| 7.927 | 7.350 | 8.768 | 6.846 | 9.095 | 11.446 |
| **7.927** | **7.350** | **7.849** | **6.278** | **9.095** | **9.607** |
| -1.129 | -2.720 | -1.662 | **0.000** | -2.107 | -3.829 |
| -1.129 | -2.720 | -3.013 | -0.297 | -2.107 | -3.917 |
| -1.129 | -2.930 | -3.803 | -0.297 | -4.987 | -3.917 |
| -1.129 | -2.930 | -5.013 | -1.781 | -5.136 | -5.477 |
| -4.425 | -6.950 | -5.720 | -1.781 | -6.592 | -5.477 |
| -4.425 | -6.950 | -6.664 | -7.235 | -6.592 | -5.877 |
| -15.935 | -14.440 | -14.065 | -14.646 | -14.231 | -14.286 |
| -15.935 | -14.440 | -14.854 | -16.857 | -14.231 | -14.286 |

## 4. Discussions

Given the excellent agreement between our results and experiment, these discussions are limited to explaining the reasons our findings are different from those from many, previous DFT calculations. Our strict adherence to the BZW-EF method is the reason our results have full physical content. For this content, it is necessary that the absolute minima of the occupied energies, i.e., the ground state, be reached – as per the variational principle of DFT also called the second theorem of Hohenberg and Kohn. The following quote from Hohenberg and Kohn [53] makes the point: "*It is well known that for a system of N particles, the energy functional of Ψ', Ev[Ψ'] =(Ψ',VΨ') + (Ψ',(T+U) Ψ') has a minimum at the correct ground state Ψ, relative to arbitrary variations of Ψ' in which the total number of particle is kept constant*". In the literature, this statement seems to have been understood only in terms of variations of linear expansion coefficients in an LCAO expansion, pursuant to iterations, using a single basis set. Numerous results of Bagayoko's group [42-50] have shown that not to be a correct view, given that iterations cannot correct for a serious deficiency of the basis set in size, angular symmetry, radial functions, and other factors such as vastly different oxidation states for w-AlN.

Most previous DFT calculations utilized a single basis set to perform self-consistent calculations. The resulting, self-consistent findings may not necessarily represent a DFT description of the concerned materials, inasmuch as DFT requires either the use of the correct charge density (i.e., correct wave function) or the verified attainment of the minimum of the above energy functional. As underscored by Ekuma et al. [49] and thoroughly explained and illustrated by Franklin et al.,[50] the linear variation to self-consistency is incapable of making up for a serious deficiency of a trial basis set. Additionally, upon reaching the absolute minima of the occupied energies, the BZW-EF method invokes the Rayleigh theorem to avoid destroying the physical content of the low energy conduction bands by using basis sets much larger than the optimal one. Such basis sets, as per the Rayleigh theorem, lower some unoccupied energies. We recall that these larger basis sets do not lower any occupied energies; the lowering of unoccupied ones is a direct consequence of the Rayleigh theorem and not the manifestation of a physical interaction. **Table V** below shows of the range of band gap values from some previous ab-initio LDA and GGA calculations, in Table I, along with the experimental findings, and our calculated gap.

**Table V.** A comparison of our result with experimental ones and with those previous ab-initio DFT calculations (with LDA and GGA potentials without adjustments) in Table 1.

| Previous LDA Calculations | Previous GGA Calculations | This work: LDA BZW-EF | Experiment |
|---|---|---|---|
| 3.9 to 4.78eV | 3.96 to 4.22 eV | **6.28** eV | 6.28 eV [5K], 6.2 eV [300K] and 6.2 $\pm$ 0.2 eV |

## 5. Conclusion

We utilized the Bagayoko, Zhao, and Williams (BZW) method, as enhanced by Ekuma and Franklin (BZW-EF), to implement the linear combination of atomic orbitals (LCAO) in ab-initio self-consistent calculations of electronic, transport, and related properties of w-AlN. We explained the reasons our findings provide a true DFT description of w-AlN. The successive calculations of the method lead to the ground state energy, i.e., the minima of the occupied energies. By invoking the Rayleigh theorem, the method avoids the destruction of the physical content of low laying unoccupied energies due to a mathematical artifact. The net result is that both the occupied energies and the low laying unoccupied ones have a full physical content. This content partly explains the excellent agreement between our calculated band gap of 6.28 eV and a measured one of the same value. Further, comparisons with some x-ray spectroscopic findings show that our partial densities of states agree qualitatively (hybridization) and quantitatively (location of peaks) with measured features of the bands. Upon the confirmation of our calculated electron effective masses by measurements, they will constitute further indication of the accuracy of our calculations – as these masses are measures of the curvatures of the concerned bands. Our results for w-AlN point to the capability of LDA BZW-EF calculations to describe accurately electronic and related properties of semi-conductors. Hence, DFT BZW-EF calculations can inform and guide the design and fabrication of semiconductor based devices.

**Acknowledgments**

This work was funded in part by the National Science Foundation (NSF) and the Louisiana Board of Regents, through LASiGMA [Award Nos. EPS- 1003897, NSF (2010-15)-RII-SUBR] and NSF HRD-1002541, the US Department of Energy – National, Nuclear Security Administration (NNSA) (Award No. DE-NA0001861), LaSPACE, and LONI-SUBR. IHN acknowledges the generous support of Ebonyi State Government, Nigeria.